\documentclass[11pt,showpacs,preprintnumbers,amsmath,amssymb]{article}
\usepackage{graphicx}
\usepackage{epsf}
\usepackage{a4wide}

\newcommand{\sect}[1]{ \section{#1} \setcounter{equation}{0} }

\newcommand{\omu}{\overline{\mu}}
\newcommand{\om}{\overline{m}}
\newcommand{\og}{\overline{g}}
\newcommand{\oD}{\overline{\Delta}}
\newcommand{\lms}{\Lambda_{\overline{\mbox{\footnotesize{MS}}}}}
\newcommand{\ms}{\overline{\mbox{MS}}}
\newcommand{\B}{\beta}
\newcommand{\C}{\gamma}
\newcommand{\obeta}{\overline{\beta}}
\newcommand{\ogamma}{\overline{\gamma}}
\newcommand{\okappa}{\overline{\kappa}}

\newcommand{\p}{\partial}
\newcommand{\f}{\frac}
\newcommand{\wD}{\widetilde{\Delta}}
\begin{document}

\title{\textbf{A determination of $\left\langle A_\mu^{2}\right\rangle$ and the
non-perturbative vacuum energy of Yang-Mills theory in the Landau
gauge}}
\author{D. Dudal\thanks{%
Research Assistant of the Fund for Scientific Research-Flanders,
Belgium.} \thanks{david.dudal@rug.ac.be} \ and H.
Verschelde\thanks{henri.verschelde@rug.ac.be} \\
{\small {\textit{Ghent University }}}\\
{\small {\textit{Department of Mathematical Physics and Astronomy }}}\\
{\small {\textit{Krijgslaan 281-S9}}}\\
{\small {\textit{B-9000 Gent, Belgium}}} \and R.E.
Browne\thanks{rebrowne@amtp.liv.ac.uk} \ and J.A.
Gracey\thanks{jag@amtp.liv.ac.uk} \\
{\small {\textit{Theoretical Physics Division}}} \\
{\small {\textit{Department of Mathematical Sciences}}} \\
{\small {\textit{University of Liverpool }}} \\
{\small {\textit{P.O. Box 147, Liverpool, L69 3BX, United
Kingdom}}}}

\date{}
\maketitle

\vspace{-10cm} \hfill LTH--569 \vspace{10cm}

\begin{abstract}
We discuss the 2-point-particle-irreducible $(2PPI)$ expansion,
which sums bubble graphs to all orders, in the context of $SU(N)$
Yang-Mills theory in the Landau gauge. Using the method we
investigate the possible existence of a gluon condensate of mass
dimension two, $\left\langle A_{\mu}^{a} A_{\mu}^{a}
\right\rangle$, and the corresponding non-zero vacuum energy. This
condensate gives rise to a dynamically generated mass for the
gluon.
\end{abstract}
\maketitle
\sect{\label{sec1}Introduction.} Recently there has been growing
evidence for the existence of a condensate of mass dimension two
in $SU(N)$ Yang-Mills theory with $N$ colours. An obvious
candidate for such a condensate is $\left\langle A^a_{\mu}
A^a_{\mu}\right\rangle$. The phenomenological background of this
type of condensate can be found in
\cite{Chetyrkin:1998yr,Gubarev:2000eu,Gubarev:2000nz}. However, if
one first considers simpler models such as massless $\lambda
\phi^{4}$ theory or the Gross-Neveu model \cite{Gross:jv} and the
role played by their quartic interaction in the formation of a
(local) composite quadratic condensate and the consequent
\emph{dynamical mass generation} for the originally massless
fields \cite{Gross:jv,Verschelde:2000dz,smet}, it is clear that
the \emph{possibility} exists that the quartic gluon interaction
gives rise to a quadratic composite operator condensate in Yang
Mills theory and hence a dynamical mass for the gluons. The
formation of such a dynamical mass is strongly correlated to a
lower value of the vacuum energy. In other words the causal
perturbative Yang Mills vacuum is unstable. From this viewpoint,
mass generation in connection with gluon pairing has already been
discussed a long time ago in, for example,
\cite{Fukuda:1977mz,Fukuda:1977zp,Gusynin:1978tr,Savvidy:1977as}.
There the analogy with the BCS superconductor and its gap equation
was examined. It was shown that the zero vacuum is tachyonic in
nature and the gluons achieve a mass due to a non-trivial solution
of the gap equation. Moreover, recent work using lattice
regularized Yang Mills theory has indicated the existence of a
non-zero condensate, $\left\langle
A^a_{\mu}A^a_{\mu}\right\rangle$, \cite{Boucaud:2001st}. There the
authors invoked the operator product expansion, (OPE), on the
gluon propagator as well as on the effective coupling $\alpha_{s}$
in the Landau gauge. Their work was based on the perception that,
even in the relatively high energy region ($\sim 10$GeV), a
discrepancy existed between the expected perturbative behaviour
and the lattice results. It was shown that, within the momentum
range accessible to the OPE, that this discrepancy could be solved
with a $1/q^{2}$ power correction\footnote{The $\f{1}{q^{4}}$
power correction due to the $\langle G_{\mu\nu}^{2}\rangle$
condensate is too weak at such energies to be the cause of the
discrepancy.}. They concluded that a non-vanishing dimension two
condensate must exist. Further, the results of
\cite{Boucaud:2002nc} give some evidence that instantons might be
the mechanism behind the low-momentum contribution to condensate.
As has been argued in \cite{Gubarev:2000nz}, only the low-momentum
content of the squared vector potential is accessible with the
OPE. Moreover, they argue that there are also
\emph{short-distance} non-perturbative contributions to
$\left\langle A_\mu^{2} \right\rangle$.

It is no coincidence the Landau gauge is used for the search for a
dimension two condensate. Naively, the operator $A_\mu^{2}$ is not
gauge-invariant. Although this does not prevent the condensate
$\left\langle A_\mu^{2}\right\rangle$ showing up in gauge variant
quantities like the gluon propagator, we should instead consider
the gauge-invariant operator $(VT)^{-1}\min_{U}\int\left.
d^{4}x\left(A_\mu^{U}\right)^{2}\right.$, where $VT$ is the
space-time volume and $U$ is an arbitrary gauge transformation in
order to assign some physical meaning to the operator. Clearly
from its structure this operator is non-local and thus is
difficult to handle. However, when we impose the Landau gauge, it
reduces\footnote{Although this equality is somewhat disturbed by
Gribov ambiguities \cite{Stodolsky:2002st}. In this paper Gribov
copies are neglected since we will work in the perturbative Landau
gauge and sum a certain class of bubble diagrams in this
particular gauge. It is a pleasant feature of the Landau gauge
that $\left\langle A_\mu^{2}\right \rangle$ can be given a gauge
invariant meaning. In another gauge, the bubbles will no longer
correspond to $\left\langle A_\mu^{2}\right\rangle$ and the
correspondence with $\left\langle A_\mu^{2}\right\rangle_{\min}$
is more of academic interest.} to the local operator $A_\mu^{2}$.
Moreover, it has been shown that $\left\langle
A_\mu^{2}\right\rangle$ is (on-shell) BRST invariant
\cite{Curci:bt,Curci:1976ar,Kondo:2001nq,Kondo:2001tm}. Another
motivation for studying $\left\langle A_\mu^{2}\right\rangle$ is
the perceived connection between the gluon propagator and
confinement. (See \cite{Langfeld:2002bg} and references therein.)
More precisely, the gluon propagator exhibits an infrared
suppression, as has been reported in many lattice simulations,
\cite{Bowman:2002fe,Bonnet:2001uh,Alexandrou:2000ja} and using the
Schwinger-Dyson approach,
\cite{Alkofer:2000mz,Alkofer:2000wg,Lerche:2002ep}. A dynamical
gluon mass might serve as an indication for such a suppression. An
attempt has already been made to explain confinement by a dual
Ginzburg-Landau model or an effective string theory, in the Landau
gauge, with the help of $\left\langle A_\mu^{2}\right\rangle$
\cite{Kondo:2002xn}. The fact that $\left\langle
A_\mu^{2}\right\rangle$ might be central to confinement, is
supported by the observation that it undergoes a phase transition
due to the monopole condensation in three dimensional compact QED
\cite{Gubarev:2000eu,Gubarev:2000nz}.

From these various analyses the importance of $\left\langle
A_\mu^{2}\right\rangle$ must have become clear. Therefore, the aim
of this article is to provide some \emph{analytical} evidence that
gluons do condense. To our knowledge, \cite{Verschelde:2001ia} is
the only paper which effectively calculates $\left\langle
A_\mu^{2} \right\rangle$, without referring to lattice
regularization. In \cite{Verschelde:2001ia} the standard way of
calculating the effective potential for a particular quantity was
followed, and all the problems concerning the fact that the
considered quantity was a composite operator were elegantly
solved.

In a previous paper \cite{dudal}, we have discussed the $2PPI$
expansion for the Gross-Neveu model and found results close to the
exact values for the Gross-Neveu mass gap and the vacuum energy.
The $2PPI$ expansion does not rely on the effective action
formalism of \cite{Verschelde:2001ia}. Instead it is directly
based on the path integral and the topology of its Feynman
diagrammatical expansion. In this paper we will discuss how to
apply it to $SU(N)$ Yang-Mills theories in the Landau gauge. Of
course, it is not our aim to provide a complete picture of
$\left\langle A_\mu^{2}\right\rangle$ but rather give further
evidence for its existence since it lowers the vacuum energy.

\sect{\label{sec2}The $2PPI$ expansion.} The $SU(N)$ Yang Mills
Lagrangian in $d$-dimensional Euclidean space time is given by
\begin{equation}\label{1}
\mathcal{L}(A_{\mu},\xi,\overline{\xi}) ~=~
\frac{1}{4}G_{\mu\nu}^{a} G_{\mu\nu}^{a} ~+~
\mathcal{L}_{gauge+F.P.}
\end{equation}
where $G^a_{\mu\nu}$ is the gluon field strength, $1$ $\leq$ $a$
$\leq$ $N^{2}-1$, $\mathcal{L}_{gauge+F.P.}$ implements the Landau
gauge and its corresponding Faddeev-Popov part and $\xi$ and
$\overline{\xi}$ denote the ghosts and anti-ghost fields
respectively. Issues concerning the counterterm part of (\ref{1})
will be discussed later. First, we consider the diagrammatical
expansion for the vacuum energy which we denote by $E$. As is well
known, this is a series consisting of one particle irreducible,
$(1PI)$, diagrams. These $1PI$ diagrams can be divided into two
disjoint classes:
\begin{itemize}
\item those diagrams not falling apart into two separate pieces
when two lines meeting at the same point $x$ are cut, which we
call 2-point-particle-irreducible, ($2PPI$); (an example is given
in Fig. 1) \item those diagrams falling apart into two separate
pieces when two lines meeting at the same point $x$ are cut which
we call 2-point-particle-reducible, ($2PPR$), while $x$ is called
the $2PPR$ insertion point; (an example is given in Fig. 2).
    \end{itemize}
        \begin{figure}[t]\label{fig1}
    \begin{center}
        \scalebox{0.5}{\includegraphics{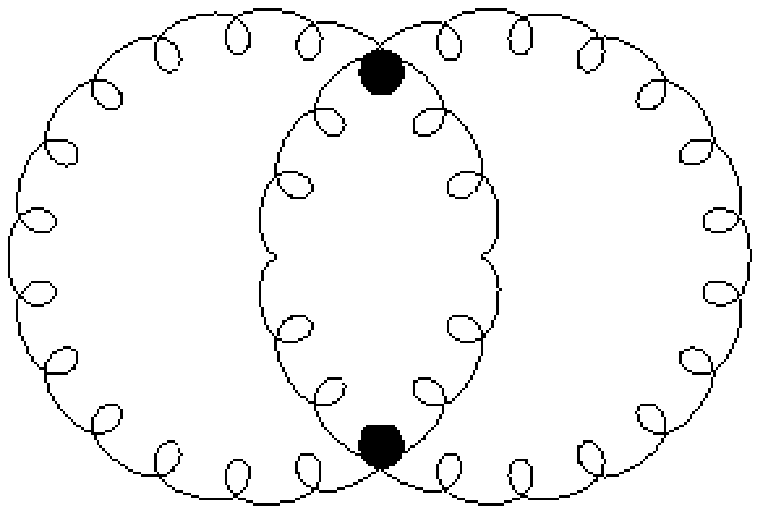}}
        \caption{A $2PPI$ vacuum bubble.}
                \scalebox{0.55}{\includegraphics{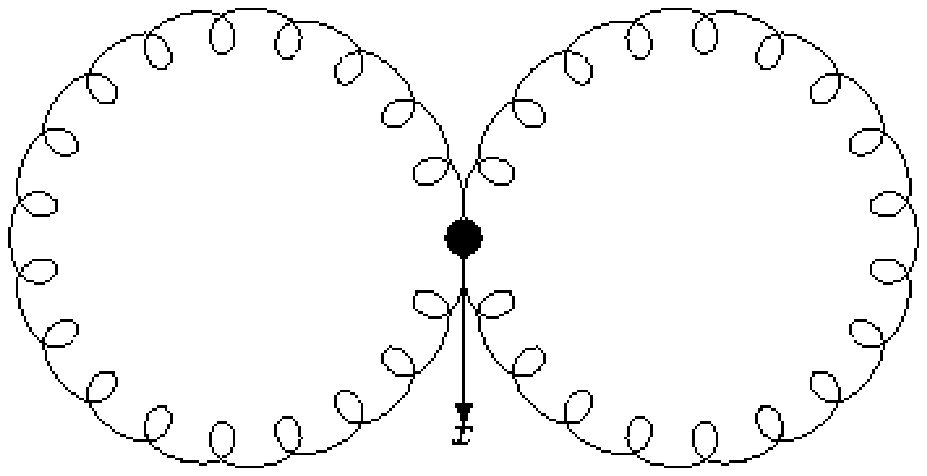}}
        \caption{A $2PPR$ vacuum bubble. $x$ is the $2PPR$ insertion point.}
    \end{center}
    \end{figure}
We may now resum this series of $2PPR$ and $2PPI$ graphs, where
the propagators are the usual massless ones, by retaining only the
$2PPI$ graphs, whereby the $2PPR$ insertions, or bubbles, are
resummed into an effective $(\textrm{mass})^{2}$
$m_{2PPI}^{2}\equiv\om^{2}$. The bubble graph gluon polarization
is then given by
\begin{eqnarray}\label{2}
\Pi^{ab}_{\mu\nu}&=& -~
\f{g^{2}}{2}\left[f_{eab}f_{ecd}\left(\left\langle
A_{\mu}^{c}A_{\nu}^{d}\right\rangle-\left\langle
A_{\mu}^{d}A_{\nu}^{c}
\right\rangle\right)+f_{eac}f_{edb}\left(\left\langle
A_{\mu}^{d}A_{\nu}^{c}\right\rangle-\left\langle
A_{\rho}^{c}A_{\rho}^{d}
\right\rangle\delta_{\mu\nu}\right)\right.\nonumber\\
&&+~\left.f_{ead}f_{ebc}\left(\delta_{\mu\nu}\left\langle
A_{\rho}^{c}A_{\rho}^{d}\right\rangle-\left\langle
A_{\mu}^{c}A_{\nu}^{d} \right\rangle\right)\right]
\end{eqnarray}
where $f_{abc}$ are the $SU(N)$ structure constants. We define the
vacuum expectation value of $A_\mu^{2}$ as
\begin{equation}\label{3}
\Delta ~=~ \left\langle A_{\mu}^{a}A_{\mu}^{a} \right\rangle ~.
\end{equation}
The global $SU(N)$ symmetry can then be used to show that
\begin{equation}\label{4}
\left\langle A_{\mu}^{a}A_{\nu}^{b}\right\rangle ~=~
\f{\delta^{ab}\delta_{\mu\nu}} {d\left(N^{2}-1\right)}\Delta
\end{equation}
Substitution of (\ref{4}) in (\ref{2}) yields
\begin{equation}\label{5}
\Pi^{ab}_{\mu\nu} ~=~ -~ g^{2}\f{N}{N^{2}-1}\f{d-1}{d}\delta^{ab}
\delta_{\mu\nu}\Delta
\end{equation}
which results in an effective mass, $\om$, running in the $2PPI$
propagators, given by
\begin{equation}\label{6}
\om^{2} ~=~ g^{2}\f{N}{N^{2}-1}\f{d-1}{d}\Delta ~.
\end{equation}
If we let $E_{2PPI}$ be the sum of the $2PPI$ vacuum bubbles,
calculated with the effective $2PPI$ propagator, then this
$E_{2PPI}$ is \emph{not} equal to the vacuum energy $E$, because
simply removing all $2PPR$ insertions is too naive. For instance,
there is a double counting problem which is already visible in the
$2PPR$ diagram of Fig. 2 where each bubble can be seen as a $2PPR$
insertion on the other one. However, we can resolve this
ambiguity. A dimensional argument results in
\begin{equation}\label{7}
E ~=~ E_{2PPI} ~+~ cg^{2}\Delta^{2}
\end{equation}
where $c\neq0$ will accomodate the double counting. To determine
the appropriate value of $c$, we use the path integral which gives
\begin{eqnarray}\label{8}
\f{\p E}{\p g^{2}} &=& -~
\f{1}{4g}f_{abc}\left\langle\left(\p_{\mu}A_{\nu}^{a}
-\p_{\nu}A_{\mu}^{a}\right)A_{\mu}^{b}A_{\nu}^{c}\right\rangle ~-~
\f{1}{2g}f_{abc}\left\langle\p_{\mu}\overline{\xi}^{a}\xi^{c}A_{\mu}^{b}
\right\rangle \nonumber \\
&& +~ \f{1}{4}f_{abc}f_{ade}\left\langle A_{\mu}^{b} A_{\nu}^{c}
A_{\mu}^{d} A_{\nu}^{e}\right\rangle ~.
\end{eqnarray}
The first two terms contribute unambiguously to the $2PPI$ part.
For the last term, we rewrite
\begin{eqnarray}\label{9}
\left\langle A_{\mu}^{b} A_{\nu}^{c} A_{\mu}^{d}
A_{\nu}^{e}\right\rangle &=& \left\langle
A_{\mu}^{b}A_{\nu}^{c}\right\rangle\left\langle A_{\mu}^{d}
A_{\nu}^{e}\right\rangle ~+~ \left\langle
A_{\mu}^{b}A_{\mu}^{d}\right\rangle
\left\langle A_{\nu}^{c}A_{\nu}^{e}\right\rangle \nonumber \\
&& +~ \left\langle A_{\mu}^{b}A_{\nu}^{e}\right\rangle\left\langle
A_{\nu}^{c} A_{\mu}^{d}\right\rangle+\left\langle A_{\mu}^{b}
A_{\nu}^{c} A_{\mu}^{d} A_{\nu}^{e}\right\rangle_{2PPI} ~.
\end{eqnarray}
Using (\ref{4}) and the properties of the structure constants
$f_{abc}$, we obtain
\begin{eqnarray}\label{10}
\f{\p E}{\p g^{2}} &=& -~ \f{1}{4g}f_{abc}\left\langle
\left(\p_{\mu}A_{\nu}^{a}-\p_{\nu}A_{\mu}^{a}\right)A_{\mu}^{b}A_{\nu}^{c}
\right\rangle_{2PPI} ~-~
\f{1}{2g}f_{abc}\left\langle\p_{\mu}\overline{\xi}^{a}
\xi^{c}A_{\mu}^{b}\right\rangle_{2PPI} \nonumber\\
&& +~ \f{1}{4}f_{abc}f_{ade}\left\langle A_{\mu}^{b} A_{\nu}^{c}
A_{\mu}^{d} A_{\nu}^{e}\right\rangle_{2PPI} ~+~
\f{1}{4}\f{N}{N^{2}-1}\f{d-1}{d}\Delta^{2}
\nonumber\\
&=&\f{\p E_{2PPI}}{\p g^{2}} ~+~
\f{1}{4}\f{N}{N^{2}-1}\f{d-1}{d}\Delta^{2} ~.
\end{eqnarray}
From (\ref{7}), we derive
\begin{eqnarray}\label{11}
\f{\p E}{\p g^{2}} ~=~ \f{\p E_{2PPI}}{\p g^{2}} ~+~ \f{\p
E_{2PPI}}{\p\om^{2}} \f{\p \om^{2}}{\p g^{2}} ~+~
c\Delta^{2}+2cg^{2}\Delta\f{\p \Delta}{\p g^{2}}~.
\end{eqnarray}
Combining (\ref{6}), (\ref{10}) and (\ref{11}) gives
\begin{eqnarray}\label{12}
\f{\p E_{2PPI}}{\p
\om^{2}}\left(\f{N}{N^{2}-1}\f{d-1}{d}\Delta+g^{2}
\f{N}{N^{2}-1}\f{d-1}{d}\f{\p \Delta}{\p g^{2}}\right) ~=~
\f{1}{4}\f{N}{N^{2}-1}\f{d-1}{d}\Delta^{2} ~.
\end{eqnarray}
Then a simple diagrammatical argument gives
\begin{eqnarray}\label{13}
    \f{\p E_{2PPI}}{\p \om^{2}} ~=~ \f{\Delta}{2} ~.
\end{eqnarray}
which is a local gap equation, summing the bubble graphs into
$\om^{2}$. Using this together with (\ref{12}) finally gives
\begin{eqnarray}\label{14}
c ~=~ -~ \f{1}{4}\f{N}{N^{2}-1}\f{d-1}{d} ~.
\end{eqnarray}
It is easy to show that the following equivalence holds
\begin{equation}\label{16}
\f{\p E_{2PPI}}{\p \om^{2}} ~=~ \f{\Delta}{2} ~\Leftrightarrow~
\f{\p E}{\p \om^{2}} ~=~ 0 ~.
\end{equation}
To summarize, we have summed the bubble insertions into an
effective $(\textrm{mass})^{2}$, $\om^{2}$. The vacuum energy is
expressed by
\begin{eqnarray}\label{15}
E ~=~ E_{2PPI} ~-~ \f{g^{2}}{4}\f{N}{N^{2}-1}\f{d-1}{d}\Delta^{2}
~.
\end{eqnarray}
We stress the fact that (\ref{15}) is \emph{only meaningful if}
the gap equation (\ref{16}) is satisfied. This means we cannot
consider $\om$ or $\Delta$ as a real variable on which $E$
depends. It is a quantity which has to obey its gap equation,
otherwise the $2PPI$ expansion loses its validity. This also means
that $E(\om)$, or equivalently $E(\Delta)$, is not a function
depending on $\om$ ($\Delta$), in contrast\footnote{$V(\varphi)$
also makes sense if $\frac{dV}{d\varphi}\neq0$.} with the usual
concept of an effective potential $V(\varphi)$ which is a function
of the constant field $\varphi$.

In order to ensure the usefulness of the $2PPI$ formalism for
actual calculations, we should prove it can be fully renormalized
with the counterterms available from the original (bare)
Lagrangian, (\ref{1}). However, it is sufficient to say that all
our derived formulae remain valid and are finite when the
counterterms are included. This also implies the $2PPI$ mass $\om$
is renormalized and gives rise to a finite, physical
mass\footnote{$m_{\mbox{\footnotesize{\textrm{phys}}}}$ is the
pole of the gluon propagator.},
$m_{\mbox{\footnotesize{\textrm{phys}}}}$. Furthermore, no new
counterterms are needed to remove the vacuum energy divergences.
The renormalizability of the $2PPI$ expansion has been discussed
in detail in \cite{dudal} in the case of the Gross-Neveu model.
Since the arguments for Yang Mills theory are completely
analogous, we refer to \cite{dudal} for technical details
concerning the renormalization.

Another point worth emphasising here, is the renormalization group
equation, (RGE), for $E$. The first diagram of $E_{2PPI}$ is given
by the O-bubble. Using the $\overline{\mbox{MS}}$ renormalization
scheme in dimensional regularization in $d$ dimensions, we find
\begin{eqnarray}\label{17}
E ~=~
\f{3}{4}\f{N^{2}-1}{16\pi^{2}}\om^{4}\left(\ln\f{\om^2}{\omu^{2}}
-\f{5}{6}\right) ~-~
\f{1}{4\og^{2}}\f{d}{d-1}\f{N^{2}-1}{N}\om^{4} ~.
\end{eqnarray}
Since $E$ is a physical quantity, it should not depend on the
subtraction scale $\omu$. This is expressed by the RGE
\begin{eqnarray}\label{18}
\omu\frac{d E}{d \omu} ~=~ \left(\omu\frac{\partial}{\partial
\omu} ~+~ \obeta(\og^{2})\frac{\partial}{\partial \og^{2}} ~+~
\okappa(\og^{2})\om^{2} \frac{\partial}{\partial \om^{2}}\right)E
~=~ 0
\end{eqnarray}
where $\obeta(\og^{2})$ governs the scaling behaviour of the
coupling constant
\begin{eqnarray}\label{19}
\obeta(\og^{2}) ~=~ \omu\f{\p \og^{2}}{\p \omu} ~=~ -~ 2\left(
\beta_{0}\og^{4} ~+~ \beta_{1}\og^{6} ~+~ \obeta_{2}\og^{8} ~+~
\cdots\right)
\end{eqnarray}
and $\okappa(\og^{2})$ is the anomalous dimension of $\om^{2}$
\begin{eqnarray}\label{20}
\omu\frac{\partial \om^{2}}{\partial \omu}
&=&\left(\frac{\obeta(\og^{2})}{\og^{2}} ~+~ \ogamma_{A_\mu^{2}}
(\og^{2})\right) \om^{2} ~\equiv~ \okappa(\og^{2})\om^{2}\\
\ogamma_{A_\mu^{2}}(\og^{2})&=&\f{\omu}{\oD}\f{\p \oD}{\p \omu}
~=~ \gamma_{0}\og^{2} ~+~ \ogamma_{1}\og^{4} ~+~
\ogamma_{2}\og^{6} ~+~ \cdots ~.
\end{eqnarray}
The coefficients can be found in \cite{gross,Gracey:2002yt} for
$\obeta$ and in
\cite{Verschelde:2001ia,Gracey:2002yt,Dudal:2002pq} for
$\ogamma_{A_\mu^{2}}$,
\begin{eqnarray}
\label{21a}
\beta_{0}&=&\f{11}{3}\left(\f{N}{16\pi^{2}}\right)\hspace{1cm}
\beta_{1} ~=~
\f{34}{3}\left(\f{N}{16\pi^{2}}\right)^{2}\hspace{1cm}
\obeta_{2} ~=~ \f{2857}{54}\left(\f{N}{16\pi^{2}}\right)^{3}\\
\label{21b}\gamma_{0}&=&\f{35}{6}\left(\f{N}{16\pi^{2}}\right)\hspace{1cm}
\ogamma_{1} ~=~
\f{449}{24}\left(\f{N}{16\pi^{2}}\right)^{2}\hspace{1cm}
\ogamma_{2} ~=~
\left[\f{75607}{864}-\f{9\zeta(3)}{16}\right]\left(
\f{N}{16\pi^{2}}\right)^{3} ~. \nonumber \\
\end{eqnarray}
When we combine all this information and determine $\omu\f{d E}{d
\omu}$ up to lowest order in $\og^{2}$, we find
\begin{equation}\label{22}
\omu\f{dE}{d\omu} ~\neq~ 0 ~.
\end{equation}
Apparently, it seems that $E$ does not obey its RGE. However, this
is not a contradiction because of the demand that the gap equation
(\ref{16}) \emph{must} be satisfied. The gap equation implies that
$\ln\f{\om^{2}}{\omu^{2}}\propto\f{1}{\og^{2}}+\textrm{constants}$.
The consequence is that all leading logarithms contain terms of
order unity. Hence, we cannot show that the RGE for $E$ is obeyed
order by order. The same phenomenon extends to higher orders. In
other words, knowledge of $\omu\f{dE}{d\omu}$ up to a certain
order $n$, would require knowledge of \emph{all} leading and
subleading log terms to order $n$, to show explicitly that
$\omu\f{dE}{d\omu}$~$=$~$0$. We must therefore be careful not to
interpret the non-vanishing of the RGE as a reason to introduce a
``non-perturbative'' $\beta$-function, as is sometimes done,
\cite{Yang:2002ur}.

\sect{\label{sec3} Results.}
\begin{figure}[t]\label{fig3}
    \begin{center}
        \scalebox{0.75}{\includegraphics{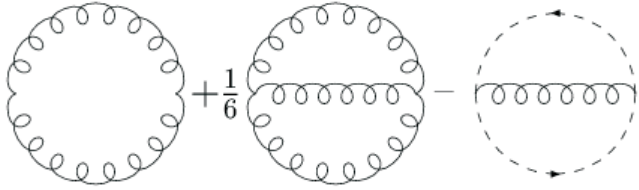}}
        \caption{The first diagrams contributing to $E_{2PPI}$}
    \end{center}
\end{figure}
Up to 2-loop order in the $2PPI$ expansion (see Fig. 3), we find
in the $\ms$ scheme
\begin{eqnarray}\label{23}
E(\oD)&=& -~ \frac{3}{16}\frac{\og^{2}N}{N^{2}-1}\oD^{2} ~+~
\frac{27}{64}\frac{\og^{4}N^{2}}{16\pi^{2}}\frac{\oD^{2}}{N^{2}-1}
\left(\ln\frac{\frac{3}{4}\frac{\og^{2}N}{N^{2}-1}\oD}{\omu^{2}}
-\frac{5}{6}\right) ~+~
\frac{9}{16}\frac{\og^{6}N^{3}}{\left(16\pi^{2}
\right)^{2}} \frac{\oD^{2}}{N^{2}-1}\nonumber\\
&& \times ~
\left[-\frac{31}{2}\left(\ln\frac{\frac{3}{4}\frac{\og^{2}N}
{N^{2}-1}\oD}{\omu^{2}}\right)^{2} ~+~
\frac{259}{8}\ln\frac{\frac{3}{4}
\frac{\og^{2}N}{N^{2}-1}\oD}{\omu^{2}} ~-~ \frac{1043}{32} ~+~
\f{891}{16}s_{2} ~-~ \frac{63}{8}\zeta(2)\right]\nonumber\\
\end{eqnarray}
where $\zeta(n)$ is the Riemann zeta function,
\begin{equation}\label{cl}
s_{2} ~=~
\f{4}{9\sqrt{3}}\mathcal{C}\ell_{2}\left({\f{\pi}{3}}\right) ~
\approx ~ 0.2604341
\end{equation}
and $\mathcal{C}\ell_2(x)$ is the Clausen function. We have
computed the relevant two loop vacuum bubble diagrams to the
finite part using the massive gluon and massless ghost propagators
which are respectively
\begin{equation}
-~ \frac{1}{p^{2}+\om^{2}} \left[ \delta_{\mu\nu} ~-~ \frac{p_\mu
p_\nu}{p^{2}} \right] ~~~~~ \mbox{and} ~~~~~ \frac{1}{p^{2}}
\end{equation}
in the Landau gauge. The expressions for the general massive and
massless two loop bubble integrals were derived from the results
of \cite{Ford:1992pn} and implemented in the symbolic manipulation
language {\sc Form}, \cite{form}. It is easy to check that solving
the gap equation $\f{\partial E}{\partial \oD}$~$=$~$0$, with
$\omu^{2}$ set equal to $\frac{3}{4}\frac{\og^{2}N}{N^{2}-1}\oD$
to kill potentially large logarithms, gives no solution at 1-loop
or 2-loop, apart from the trivial one $\oD$~$=$~$0$. This does not
imply $\left\langle A_\mu^{2}\right\rangle$ does not exist but
that the $\ms$ scheme might not be the best choice for the $2PPI$
expansion. To address this we first remove the freedom existing in
how the mass parameter $\Delta$ is renormalized by replacing
$\Delta$ by a renormalization scheme and scale independent
quantity $\wD$. This can be accommodated by\footnote{Barred
quantities refer to the $\ms$ scheme, otherwise any other massless
renormalization scheme is meant.}
\begin{equation}\label{24}
\wD ~=~ \overline{f}\left(\og^{2}\right)\oD
\end{equation}
with
\begin{equation}\label{25}
\omu\frac{\p \overline{f}}{\p \omu} ~=~ -~
\ogamma_{A_\mu^{2}}\left(\og^{2} \right) \overline{f} ~.
\end{equation}
A change in massless renormalization scheme corresponds to
relations of the form
\begin{eqnarray}
\label{26a}\og^{2}&=&g^{2}\left(1+b_{0}g^{2}+b_{1}g^{4}+\cdots\right)\\
\label{26b}\oD&=&\Delta\left(1+d_{0}g^{2}+d_{1}g^{4}+\cdots\right)\\
\label{26c}\overline{f}(\og^{2})&=&f(g^{2})\left(1+f_{0}g^{2}+f_{1}g^{4}
+\cdots\right) ~.
\end{eqnarray}
With these, it is easily checked that (\ref{24}) is
renormalization scheme and scale independent. The explicit
solution of (\ref{25}) reads
\begin{eqnarray}\label{27}
\overline{f}\left(\og^{2}\right)&=&\left(\og^{2}\right)^{\frac{\gamma_{0}}
{2\beta_{0}}}
\left\{1\vphantom{\frac{\C_{0}\left(\frac{\B_{1}^{2}}{\B_{0}^{2}}
-\frac{\obeta_{2}}{\B_{0}}\right)}{\B_{0}}}
+\frac{\og^{2}}{2}\left(-\frac{\B_{1}\C_{0}}{\B_{0}^{2}}
+\frac{\ogamma_{1}}{\B_{0}}\right)+\frac{\og^{4}}{4}
\left[\vphantom{\frac{\C_{0}\left(\frac{\B_{1}^{2}}{\B_{0}^{2}}
-\frac{\obeta_{2}}{\B_{0}}\right)}{\B_{0}}}
\frac{1}{2}\left(-\frac{\B_{1}\C_{0}}{\B_{0}^{2}}
+\frac{\ogamma_{1}}{\B_{0}}\right)^{2}\right.\right.\nonumber\\
&& +~ \left.\left.\frac{\C_{0}\left(\frac{\B_{1}^{2}}{\B_{0}^{2}}
-\frac{\obeta_{2}}{\B_{0}}\right)}{\B_{0}}-\frac{\B_{1}\ogamma_{1}}{\B_{0}^{2}}
+\frac{\ogamma_{2}}{\B_{0}}\right]+\mathcal{O}\left(\og^{6}\right)\right\}
~.
\end{eqnarray}
Since the gap equation is still a series expansion in
$\f{g^{2}N}{16\pi^{2}}$ and we hope to find (at least
qualitatively) acceptable results, $\f{g^{2}N}{16\pi^{2}}$ should
be small. We will therefore choose to renormalize the coupling
constant in such a scheme so that $E$ is of the form
\begin{equation}\label{e1}
E ~=~
\f{3}{16}\left(g^{2}\right)^{1-\f{\C_{0}}{\B_{0}}}\f{N}{N^{2}-1}\wD^{2}
\left(-1+\f{g^{2}N}{16\pi^{2}}E_{1}^{1} L
+\left(\f{g^{2}N}{16\pi^{2}}\right)^{2}\left(E_{2}^{1}L +
E_{2}^{2}L^{2}\right)+\ldots\right)
\end{equation}
where
\begin{equation}\label{e1bis}
L ~=~
\ln\frac{\frac{3}{4}\left(g^{2}\right)^{1-\frac{\gamma_{0}}{2\beta_{0}}}
\frac{N}{N^{2}-1}\wD}{\omu^{2}} ~.
\end{equation}
Otherwise, we remove all the terms of the form
$\left(\f{g^{2}N}{16\pi^{2}}\right)^{n}\times\textrm{constant}$,
and only keep the terms that contain a power of the logarithm $L$.
This is always possible by calculating the $\ms$ value of $E$ as
in (\ref{23}) and using (\ref{26a}) to change the coupling
constant renormalization by a suitable choice for the coefficients
$b_{i}$. In other words the 1-loop $\ms$ contribution to $E$
allows one to determine $b_{0}$. Once $b_{0}$ is fixed, the 2-loop
$\ms$ contribution to $E$ can be used to fix $b_{1}$, and so on
for the higher order contributions. We note that the gap equation
(\ref{16}) is translated into $\frac{\p E}{\p\wD}$~$=$~$0$ since
$\om^{2}\propto\overline{\Delta}\propto\wD$. In this gap equation,
we will set
$\omu^{2}$~$=$~$\frac{3}{4}\left(g^{2}\right)^{1-\frac{\gamma_{0}}
{2\beta_{0}}}\frac{N}{N^{2}-1}\wD$ so that all logarithms vanish.
In other words $L$~$=$~$0$. Notice that one cannot set
$\omu^{2}$~$=$~$\frac{3}{4}\left(g^{2}\right)^{1-\frac{\gamma_{0}}
{2\beta_{0}}}\frac{N}{N^{2}-1}\wD$ in the expression (\ref{e1})
and use the RGE for $E$ to sum the logarithms when
$\frac{dE}{d\wD}=0$ is solved. As already explained in the
previous section, the RGE for $E$ is not obeyed order by order,
see also \cite{dudal}. Once a solution $\wD_{*}$ of the gap
equation is found, then we will always have
\begin{equation}\label{e1tris}
E_{\mbox{\footnotesize{\textrm{vac}}}} ~=~ -~ \f{3}{16}\left(g^{2}
\right)^{1-\f{\C_{0}} {\B_{0}}}\f{N}{N^{2}-1}\wD_{*}^{2} ~.
\end{equation}
If the constructed value for $\f{g^{2}N}{16\pi^{2}}$ is small
enough, then we can trust, at least qualitatively, the results we
will obtain. Now we are ready to rewrite (\ref{23}) in terms of
$\wD$. After a little algebra, one finds
\begin{eqnarray}\label{29}
E&=&\frac{3}{16}\left(g^{2}\right)^{1-\f{\C_{0}}{\B_{0}}}\frac{N}{N^{2}-1}
\wD^{2}\left\{-1+\left[\f{9N}{64\pi^{2}}\left(-\f{5}{6}+L\right)-2c_{1}
-c_{4}\right]g^{2} \right.\nonumber\\
&& +~
\left.\left[\f{3N^{2}}{256\pi^{4}}\left(c_{3}+\f{259}{8}L-\f{31}{2}L^{2}
\right)-2b_{0}c_{1}-c_{1}^{2}-2c_{2}+\left(\f{9N}{64\pi^{2}}\left(-\f{5}{6}
+L\right)-2c_{1}\right)c_{4} \right.\right.\nonumber\\
&& -~ c_5 +
\left.\left.\f{9N}{64\pi^{2}}\left(\left(-\f{5}{6}+L\right)b_{0}
+c_{1}
+2\left(-\f{5}{6}+L\right)c_{1}+b_{0}\left(1-\f{\C_{0}}{2\B_{0}}
\right)\right) \right]g^{4}+\mathcal{O}\left(g^{6}\right)\right\} \nonumber \\
\end{eqnarray}
with
\begin{eqnarray}
\label{29bisa}c_{1}&=&\frac{1}{2}\left(\frac{\B_{1}\C_{0}}{\B_{0}^{2}}
-\frac{\ogamma_{1}}{\B_{0}}\right)\\
\label{28bisb}c_{2}&=&\frac{1}{8}\left(-\frac{\B_{1}\C_{0}}{\B_{0}^{2}}
+\frac{\ogamma_{1}}{\B_{0}}\right)^{2}-\frac{1}{4}\left(\frac{\C_{0}\left(
\frac{\B_{1}^{2}}{\B_{0}^{2}}
-\frac{\obeta_{2}}{\B_{0}}\right)}{\B_{0}}\right)
+\frac{1}{4}\left(\frac{\B_{1}\ogamma_{1}}{\B_{0}^{2}}-\frac{\ogamma_{2}}
{\B_{0}}\right)\\
\label{29bisd}c_{3}&=&-~\frac{1043}{32}-\frac{63}{8}\zeta(2)+\f{891}{16}s_{2}\\
\label{29bise}c_{4}&=&b_{0}\left(1-\f{\C_{0}}{\B_{0}}\right)\\
\label{29bisf}c_{5}&=&b_{1}\left(1-\f{\C_{0}}{\B_{0}}\right)-b_{0}^{2}
\f{\C_{0}}{2\B_{0}}\left(1-\f{\C_{0}}{\B_{0}}\right) ~.
\end{eqnarray}
Next, we determine $b_{0}$ and $b_{1}$ so that (\ref{29}) reduces
to
\begin{eqnarray}\label{e2}
E&=&\frac{3}{16}\left(g^{2}\right)^{1-\f{\C_{0}}{\B_{0}}}\wD^{2}
\frac{N}{N^{2}-1} \left\{-1+g^{2}\f{9N}{64\pi^{2}}L
+g^{4}\left[\f{3N^{2}}{256\pi^{4}}\left(\f{259}{8}L-\f{31}{2}L^{2}
\right)\right.\right.\nonumber\\
&& +~
\left.\left.\f{9N}{64\pi^{2}}\left(b_{0}+c_{4}+2c_{1}\right)L\right]
+\mathcal{O}\left(g^{6}\right)\right\} ~.
\end{eqnarray}
We find that $b_{0}$ is
\begin{equation}\label{e3}
b_{0} ~=~ \f{409}{2288}\f{N}{\pi^{2}} ~.
\end{equation}
We do not list the value for $b_{1}$ since it is no longer
required. From the $\beta$-function we find the two loop
expression for the coupling constant is
\begin{eqnarray}\label{30}
{g}^{2}(\omu)&=&\frac{1}{\beta_{0}\ln\frac{\omu^{2}}{\Lambda^{2}}}
-\frac{\beta_{1}}{\beta_{0}}
\frac{\ln\ln\frac{\omu^{2}}{\Lambda^{2}}}
{\beta_{0}^{2}\ln^{2}\frac{\omu^{2}}{\Lambda^{2}}}
\end{eqnarray}
where $\Lambda$ is the scale parameter of the corresponding
massless renormalization scheme. We will express everything in
terms of the $\ms$ scale parameter $\lms$. In \cite{Celmaster:km},
it was shown that
\begin{equation}\label{32}
\Lambda ~=~ \lms e^{-\frac{b_{0}}{2\beta_{0}}} ~.
\end{equation}
We will also derive a value for the
$\left\langle\f{\alpha_{s}}{\pi} G_{\mu\nu}^{2}\right\rangle$
condensate from the trace anomaly
\begin{equation}
\Theta_{\mu\mu} ~=~
\f{\beta(g)}{2g}\left(G^{a}_{\rho\sigma}\right)^{2} ~.
\end{equation}
This anomaly allows us to deduce for $N$~$=$~$3$ the following
relation between the vacuum energy and the gluon condensate
\begin{equation}\label{ano}
\left\langle\f{\alpha_{s}}{\pi} G_{\mu\nu}^{2}\right\rangle ~=~ -~
\f{32}{11}E_{\mbox{\footnotesize{\textrm{vac}}}} ~.
\end{equation}
At 1-loop order, the results for $N$~$=$~$3$ are
\begin{eqnarray}
\label{res1a}\left.\f{g^{2}N}{16\pi^{2}}
\right|_{\mbox{\footnotesize{\textrm{1-loop}}}}
&=&\f{8}{9} \\
\label{res1b}\left.\sqrt{\wD}\right|_{\mbox{\footnotesize{\textrm{1-loop}}}}
&\approx&1.004\lms\approx233\textrm{MeV}  \\
\label{res1c}\left.E_{\mbox{\footnotesize{\textrm{vac}}}}
\right|_{\mbox{\footnotesize{\textrm{1-loop}}}}
&\approx& -0.0074\lms^{4} ~ \approx ~ -~ 0.00002\textrm{GeV}^{4} \\
\label{res1D} \left. \left\langle\f{\alpha_{s}}{\pi}
G_{\mu\nu}^{2}\right\rangle
\right|_{\mbox{\footnotesize{\textrm{1-loop}}}} &\approx&
0.02\lms^{4} ~ \approx ~ 0.00007\textrm{GeV}^{4}
\end{eqnarray}
while the scale parameter $\omu^{2}\approx(184\textrm{MeV})^{2}$.
We have used $\lms$~$\approx$~$233\textrm{MeV}$ which was the
value reported in \cite{Boucaud:2001st}. We see that the 1-loop
expansion parameter is quite large and we conclude that we should
go to the next order where the situation is improved. We find
\begin{eqnarray}
\label{res2a}\left.\f{g^{2}N}{16\pi^{2}}
\right|_{\mbox{\footnotesize{\textrm{2-loop}}}}
&\approx&0.131  \\
\label{res2b}\left.\sqrt{\wD}
\right|_{\mbox{\footnotesize{\textrm{2-loop}}}}
&\approx&2.3\lms ~\approx ~ 536 \textrm{MeV} \\
\label{res2c}\left.E_{\mbox{\footnotesize{\textrm{vac}}}}
\right|_{\mbox{\footnotesize{\textrm{2-loop}}}}
&\approx& -0.63\lms^{4} ~\approx ~- ~ 0.002\textrm{GeV}^{4} \\
\label{res2d} \left. \left\langle\f{\alpha_{s}}{\pi}
G_{\mu\nu}^{2}\right\rangle
\right|_{\mbox{\footnotesize{\textrm{2-loop}}}} &\approx&
1.84\lms^{4} ~ \approx ~ 0.005\textrm{GeV}^{4} ~.
\end{eqnarray}
with $\omu^{2}\approx(347\textrm{MeV})^{2}$. Although there is a
sizeable difference between 1-loop and 2-loop results, the
relative smallness of the 2-loop expansion parameter, indicates
that the 2-loop values are qualitatively trustworthy. It is well
known that in order to find reliable perturbative results, one
must go beyond 1-loop, and even beyond 2-loop approximations.
Therefore, one should not attach a firm quantitative meaning to
the numerical values. Let us compare our results with what was
found elsewhere with different methods. A combined lattice fit
resulted in $\sqrt{\left\langle
A_\mu^{2}\right\rangle}$~$\approx$~$1.64\textrm{GeV}$,
\cite{Boucaud:2001st}. We cannot really compare this with our
result for $\sqrt{\wD}$, since the lattice value was obtained with
the OPE at a scale $\mu$~$=$~$10\textrm{GeV}$ in a specific
renormalization scheme (MOM). However, it is satisfactory that
(\ref{res2b}) is at least of the same order of magnitude. More
interesting is the comparison with what was found in
\cite{Verschelde:2001ia} with the local composite operator
formalism for $\left\langle A_\mu^{2}\right\rangle$. In the $\ms$
scheme at 2-loop order, it was found that
$\f{\og^{2}N}{16\pi^{2}}\approx 0.141247$ while
$E_{\mbox{\footnotesize{\textrm{vac}}}}$~$\approx$~$-\,0.789\lms^{4}$
which is in quite good agreement with our results. An estimate of
the tree level gluon mass of $\sim500\textrm{MeV}$  was also given
in \cite{Verschelde:2001ia} which compares well with the lattice
value of $\sim600\textrm{MeV}$ of \cite{Alexandrou:2001fh,
Langfeld:2001cz}. With the $2PPI$ method, one does not really have
the concept of a tree level mass. Instead, the determination of
$m_{\mbox{\footnotesize{\textrm{phys}}}}$ would need the
calculation of the highly non-trivial 2-loop $2PPI$ mass
renormalization graphs which is beyond the scope of this article.

In conclusion we note that the perturbative Yang-Mills vacuum is
unstable and lowers its value through a non-perturbative mass
dimension two gluon condensate $\left\langle
A_\mu^{2}\right\rangle$. We have omitted quark contributions in
our analysis but it is straighforward to extend the $2PPI$
expansion to QCD with quarks included. Indeed an idea of the
effect they have could be gained by an extension of
\cite{Verschelde:2001ia}.

\vspace{1cm} \noindent {\bf Acknowledgement.} This work was
supported in part by {\sc Pparc} through a research studentship,
(REB).

\end{document}